\def\diff{\mathrm{d}}

\def\a{\mathrm{a}}
\def\b{\mathrm{b}}
\def\c{\mathrm{c}}
\def\d{\mathrm{d}}
\def\A{\mathrm{A}}
\def\B{\mathrm{B}}
\def\C{\mathrm{C}}
\def\D{\mathrm{D}}

\def\vecr{\mathbf{r}}
\def\vecp{\mathbf{p}}
\def\Ntest{N_{\textrm{test}}}
\def\bbsty#1#2#3{#1 (#3) #2}	

\documentclass[final,5p,times,twocolumn, psfig,amsmath,amssymb]{elsarticle}
\usepackage{graphicx}\usepackage{epsfig}
\usepackage{amsmath}\usepackage{amssymb}\usepackage{amsfonts}
\usepackage{bm}
\usepackage{exscale}
\usepackage{dcolumn}    
\usepackage{color}
\usepackage{soul}
\usepackage{ulem}
\begin{document}
%
%
\title{Bifurcations in Boltzmann-Langevin One Body dynamics for fermionic systems}
%
%
\author[1]{P.Napolitani}
\author[2]{M.Colonna}
%
%
\address[1]{IPN, CNRS/IN2P3, Universit\'e Paris-Sud 11, 91406 Orsay cedex, France}
\address[2]{INFN-LNS, Laboratori Nazionali del Sud, 95123 Catania, Italy}
%
%
\begin{abstract}
	We investigate the occurrence of bifurcations in the dynamical trajectories 
depicting central nuclear collisions at Fermi energies. 
	The quantitative description of the reaction dynamics is obtained within 
a new transport model,
based on the solution of
the Boltzmann-Langevin equation in three dimensions, 	
with a broad applicability for dissipative fermionic dynamics.

	Dilute systems formed in central collisions are shown to fluctuate 
between two energetically favourable mechanisms: 
reverting to a compact shape or rather disintegrating into 
several fragments. The latter result can be connected to the recent observation
of bimodal distributions for quantities characterising fragmentation
processes and may suggest new investigations. 

\end{abstract}

%
%
%
%
%
\maketitle
\thispagestyle{plain}
%
%
%
%
%


%
\section{Introduction}
Phase transitions are general phenomena occurring in interacting
many-body systems~\cite{Schmidt_Hock,Kim2004,Steinheimer2012,Moretto2011}. 
Over the past years, many efforts have been devoted to the identification
of new features related to finite-size effects. 
	As shown by recent thermodynamical analyses,
first-order phase transitions in finite systems are characterised by
negative specific heat and
bimodal behaviour of the distribution of the order parameter~\cite{Binder1984,Chomaz2000_2003}.
The latter physically corresponds to the simultaneous presence of different classes
of physical states for the same value of the system conditions that trigger the transition
(like the temperature, for instance).  

In particular, the appearance of phase transitions from the liquid to the vapour phases  
has been widely investigated 
in the context of the nuclear multifragmentation phenomenon~\cite{Bowman1991,Borderie2008,Moretto2011,Dagostino2000}. 
Indeed, due to the analogies between the nuclear forces and the Van-der-Waals interaction,  
the nuclear matter equation of state (EOS) foresees 
such a possibility~\cite{Jacquaman1983,Muller1995}. 
The theoretical findings cited above have stimulated 
corresponding thermodynamical analyses of the properties of the products issued
from nuclear reactions at Fermi energies. Under suitable conditions,  
a bimodal character of 
experimental observables, such as the size of the heaviest cluster
produced in each collision event~\cite{Bonnet2009}, or the
asymmetry between the charges of
the two heaviest reaction products~\cite{Pichon2006} 
has been revealed.
Many investigations have also been focused on the complex
nuclear many-body dynamics, to probe the reaction mechanisms governing
the occurrence of phase transitions~\cite{Aichelin1991,Morawetz2007,Colonna2010,Chomaz2004}. 
Within such a context, nuclear fragmentation studies 
at intermediate energies (above 50 MeV per nucleon)
have recently pointed out that bimodality could have a dynamical origin, related to 
the fragment-formation mechanism~\cite{LeFevre2009}, without
necessarily requiring the reaching of thermodynamical equilibrium. 
%

From a general point of view, 
interacting many-body systems may experience a very rich dynamics,
ranging from mechanisms dominated by one-body (mean-field) effects 
to phenomena governed by strong fluctuations and correlations. 
	In the regime of low-energy collective processes,
nuclear dynamics presents a 
rather stable character;
this is the domain where the fluctuation mechanism can be described
in the small amplitude limit, restricting to mean-field (quantum) fluctuations of 
collective observables~\cite{Abe1996,Lacroix2012}.
	This limit is exceeded when violent perturbations, like for instance 
dissipative heavy-ion collisions, bring the system beyond the one-body 
collective dynamics, with two-body nucleon collisions and correlations 
playing an important role. 
Along the compression-expansion path traced by the nuclear reaction,
fluctuations introduce the anisotropy seeds from 
which `nuclear droplets' can develop. 
More precisely,
the system may access
mechanically unstable regions of the EOS, called spinodal,
where a density rise is related to a pressure fall;
there, phase-space fluctuations are even amplified, 
leading to phase separation~\cite{Chomaz2004,Borderie2008}.
As soon as a mottling pattern stands out at low density, 
i.e. at the boundary of the phase separation, a bundle of 
bifurcations into a variety of different dynamical paths may set in.

%
\section{The Boltzmann-Langevin-One-Body model}
%
	The aim of this work is to further investigate the dynamical trajectory of
disassembling nuclear systems, seeking for features associated with phase transitions. 
	In particular,
we will explore the possible occurrence of bifurcation patterns
and bimodal behaviour in central heavy-ion reactions 
at beam energies around the multifragmentation threshold.

This study is undertaken in the framework 
of a new numerical implementation of the Boltzmann-Langevin (BL) equation, 
well suited to describe out-of-equilibrium processes, such as nuclear collisions:
the Boltzmann-Langevin-One-Body (BLOB) model.
The BL equation describes the time evolution 
of the semiclassical one-body distribution function $f({\bf r},{\bf p},t)$ in response to the 
mean-field potential, 
incorporating the effect of
fluctuations and correlations due to hard two-body 
scattering~\cite{Ayik1990,Randrup1990,Burgio1991,Colonna1998}.
%
%

%
 	Hence the distribution function $f$ evolves according to
the action of the effective Hamiltonian $H[f]$, the average Boltzmann collision integral 
$\bar{I}[f]$, and the fluctuating term $\delta I[f]$ as:
\begin{equation}
	\dot{f}	= \partial_t\,f - \left\{H[f],f\right\}
			= {\bar{I}[f]}+{\delta I[f]} \;.
\label{eq1}
\end{equation}
%
%
	This form indicates that the residual interaction, represented by the 
right-hand side
of eq.~\ref{eq1}, is expressed in terms of the one-body distribution function $f$.

	Like in standard transport approaches~\cite{Bertsch1988}, 
we sample the dynamics through
the test-particle method, under the assumption of spatial and temporal 
locality of the two-body collisional process.
$\Ntest$ particles per nucleon are employed. 	
%
	The Boltzmann-Langevin theory  
describes fluctuations of $f$ on a size scale of $h^3$, but it leaves the
shape of such a phase-space volume 
arbitrary (see the discussion in  ref.~\cite{Rizzo2008}).
	This same arbitrarity characterises other molecular-dynamics or Boltzmann-like
approaches.
	In this respect, we chose to follow the prescription of Bauer and Bertsch~\cite{Bauer1987}:
in order to solve the BL equation, they proposed to define nucleon wave packets 
by organising  test particles in phase-space agglomerates of $\Ntest$ elements.
	However, in ref.~\cite{Bauer1987} Pauli blocking was checked only 
for the centroids of the nucleon clouds: the effect of such 
approximation on the fermionic dynamics was analysed in 
refs.~\cite{Burgio1991,Chapelle1992}, where it was concluded that 
an incomplete treatment of Pauli blocking affects the mechanism of 
fluctuation development.
	The above recommendation was taken into account in 
ref.~\cite{Rizzo2008}, in treating the schematic case of nuclear 
matter in a periodic box: such approach confirmed that an accurate 
treatment of the Pauli blocking is the key for correctly describing 
the fluctuation mechanism in full phase space.
	By further improving the above approach in a full model for 
heavy-ion collisions, we built a novel numerical procedure where 
nucleon-nucleon (N-N) correlations are implemented by accurately 
treating the Pauli-blocking factors of agglomerates of $\Ntest$ 
elements of identical isospin.
	This arrangement, which simulates nucleon wave packets,
is redefined at successive time steps and locally, 
for couples of colliding agglomerates. 
	According to this rescaling and for elastic N-N collisions only, 
the average rate of change of the occupancy $f_a$ around the phase-space location 
$(\vecr_a,\vecp_a)$ at a given time takes the form:
\begin{equation}
 	\dot{f}_a(\vecr_a,\vecp_a) 
	=\;g\int\frac{\diff\vecp_b}{h^3}\,
	\int \diff\Omega\;\;
	W({\scriptstyle\A\B\leftrightarrow\C\D})\;
	F({\scriptstyle\A\B\rightarrow\C\D})
	\,,
\label{eq2}
\end{equation}
	where $g$ denotes the degeneracy, and integrations are over 
momenta $\vecp_b$ and scattering angles $\Omega$. 
The first integration argument is the symmetric transition rate from an
intermediate state $\A\B$ to a final configuration $\C\D$. 
To select the test particles defining  the nucleon wave packets $\A$ and $\B$,
we adopt the following procedure: 
We consider a sphere, centred at the position ${\bf r}_a$, with radius equal to the scattering distance, 
associated with the free elastic N-N 
cross section at Fermi energies (taken equal to 50 mb);
among all test particles inside the sphere, we pick up the $\Ntest$ closest particles 
to the elements ${a}$ and ${b}$ in momentum space, respectively. 
The final state is represented 
by ${c}\in \C$ and ${d}\in \D$. The transition rate is obtained by
averaging over all couples of test particles 
involved in the transition $\Sigma = (\A\B\!\rightarrow\!\C\D)$,
in terms of relative velocity and differential N-N cross section:
\begin{equation}
	W({\scriptstyle\A\B\leftrightarrow\C\D}) = 
	\Big\langle |v_{a}\!-\!v_{b}| \frac{\diff\sigma}{\diff\Omega} \Big\rangle_\Sigma = 
	\Big\langle W({\scriptstyle\a\b\leftrightarrow\c\d})\Big\rangle_\Sigma
	\,.
\label{eq3}
\end{equation}
	The second integration argument contains the product of occupancies of the
entire agglomerates $f_{\A\ldots\D}$ and of the associated vacancies 
$\bar{f}_{\A\ldots\D}$:
\begin{equation}
	F({\scriptstyle\A\B\rightarrow\C\D}) =
	\bar{f}_\A\bar{f}_\B f_\C f_\D\!-\!f_\A f_\B \bar{f}_\C\bar{f}_\D
	= \Big\langle\!F({\scriptstyle\a\b\rightarrow\c\d})\!\Big\rangle_\Sigma
	\,.
\label{eq4}
\end{equation}

	Rewritten in terms of test-particles, the representation of eq.~\ref{eq4} 
indicates that only the fraction of the packets which are really modified by the
scattering can significantly contribute to the transition probability, 
while overlapping volumes contribute to the Pauli-blocking factors.

	On this basis, full phase-space fluctuations are introduced in the 
equation of motion by moving simultaneously 
the test-particle agglomerates,
in analogy with the extended-TDHF procedure of including 
perturbations in the Slater configuration~\cite{Reinhard1992}.
	The scattering is decided by confronting the probability $W\times F$ with 
a random number and scanning the entire phase space in search of collision 
configurations at successive time steps. 
	Since all test particles belonging to the
agglomerates $\A$ and $\B$ can be reconsidered as starting points of
new collision processes,
the scattering probability has to be suitably rescaled, dividing it by
$\Ntest^2$.
	Once the sorting allows for a scattering to occur, modulation functions
are applied to precisely adapt the density profile of final-states 
to the available vacancy profile $\bar{f}_{\A\ldots\D}$, with the requirement of imposing 
the most compact configuration compatible with the constraint of energy 
conservation~\cite{Napolitani2012}; the resulting occupation functions of the modulated
final-state density profiles $(f_{\A\ldots\D})_{\textrm{M}}$ should approach 
unity.

	The extension of the wave packets makes necessary to pay special 
attention to scatterings close to the surface of the system, i.e. occurring across potential boundaries:
confronting the shape of the wave packet to the shape of
the surface,
the blocking factors are increased
in proportion to the spread of the nucleon packet 
outside of the boundary 
(similarly to what is done in some molecular-dynamics approaches~\cite{Aichelin1991}).

It occurs in some situations, for instance when low densities are 
attained, that the nuclear system is brought to explore regions of the phase 
diagram where it becomes unstable against density fluctuations, like the 
spinodal region. 
	The action of the BL term results in agitating the density profile 
over several wave lengths. 
	It can be proven for the proposed BL approach (through an analysis of the linear response in the mean field, see refs.~\cite{Colonna1994,Chomaz2004}) 
that the amplitude of the unstable modes grows according to the specific dispersion relation associated with the employed mean-field interaction.

\section{Application to head-on heavy-ion collisions at Fermi \mbox{energies}}
%
%
%
\begin{figure}[b!]\begin{center}
	\includegraphics[angle=0, width=1\columnwidth]{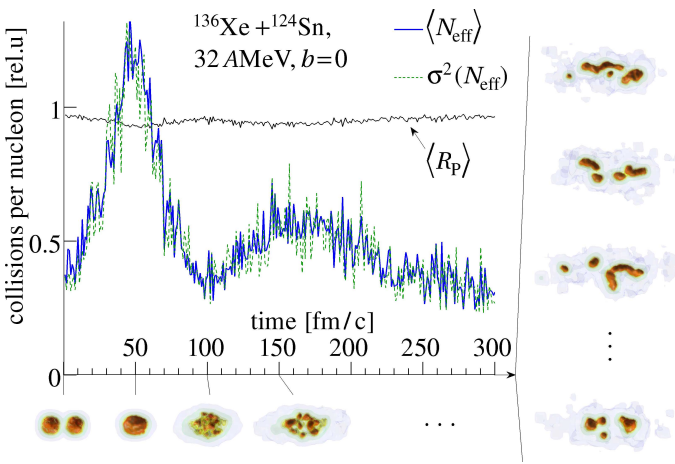}
\end{center}\caption
{
	(Color online) N-N collisions in $^{136}$Xe$+^{124}$Sn at $32\,A$MeV 
	for $b\!=\!0$:
	mean Pauli-blocking rejection rate 
	$\langle R_{\textrm{P}}\rangle$,
	and mean evolution of the number of effective collisions 
	$\langle N_{\textrm{eff}}\rangle$ which equals its 
	variance $\sigma^2(N_{\textrm{eff}})$.
	Bottom band. Time evolution of the projected density profile.
	Right band. For the same initial conditions, bifurcations lead 
	to several different fragmentation patterns at 300 fm/c.
}
\label{fig_collisions_statistics}
\end{figure}
%
In the following, the BLOB model is applied 
to 
a highly constraining 
phenomenology: the low-energy threshold for multifragmentation in
head-on heavy-ion collisions at Fermi energies.
We will also refer to results of the so-called Stochastic Mean Field (SMF) 
model~\cite{Colonna1998,Chomaz2004}, that corresponds to an approximate treatment
of the BL equation. 
In SMF, fluctuations are projected on the coordinate space and injected by agitating the spatial density profile. 
We will show that  the implementation of fluctuations in full phase space (as in BLOB) improves
significantly the reaction dynamics.

	Like in ref.~\cite{Baran2005}, for the simulation we use a soft 
equation of state with $k_{\inf}\!=\!200$ MeV;
the potential component of the symmetry energy of the EOS is
given by a linear term as a function of the density (asy-stiff).
%
%
	For numeric purposes, the transition rate has been rewritten as
$W\approx\langle|v_a\!-\!v_b|\rangle_\Sigma(\diff\sigma_{\textrm s}/\diff\Omega)$
by introducing a screened in-medium cross section $\sigma_{\textrm s}$ 
(from ref.~\cite{DanielewiczCoupland}),
corresponding to the transition $(\A\B\!\leftrightarrow\!\C\D)$ .

Fig.~\ref{fig_collisions_statistics} illustrates a study of the N-N collision 
statistics for the collision 
$^{136}$Xe$+^{124}$Sn at 32~$A$MeV,
at a central impact parameter ($b=0$).
The number of events considered is 600.
	Only a tiny fraction of the attempted N-N collisions is permitted by
Pauli-blocking factors. 
	Indeed the Pauli-blocking rejection rate $R_{\textrm{P}}$, defined
as the ratio between rejected and attempted collisions, 
is close to 
unity.
	The average number of effective N-N collisions $\langle N_{\textrm{eff}}\rangle$
reflects the average density evolution of the system: 
a maximum is encountered in correspondence with the largest 
compression and a minimum marks the subsequent expansion mechanism.
	This evolution is illustrated for one single event in the lower band of 
fig.~\ref{fig_collisions_statistics}.
	Later on, $\langle N_{\textrm{eff}}\rangle$ increases again indicating 
that the process of fragment formation sets in, and it finally levels 
off till, around $300$ fm/c, the system has stabilised into a well 
defined configuration.
	Thus fragments are formed when the system expands, according to the spinodal
decomposition scenario~\cite{Chomaz2004}. 
	The equality between the mean value and the variance ensures that 
Langevin fluctuations stand out with the correct 
amplitude~\cite{Rizzo2008}. 
	Some examples of fragmentation patterns appearing at $300$ fm/c
for the same initial conditions are collected in the right band of the figure; 
their variety already signs the presence of bifurcations.

	\subsection{Onset of fragmentation}
%
%
%
\begin{figure}[t!]\begin{center}
	\includegraphics[angle=0, width=1\columnwidth]{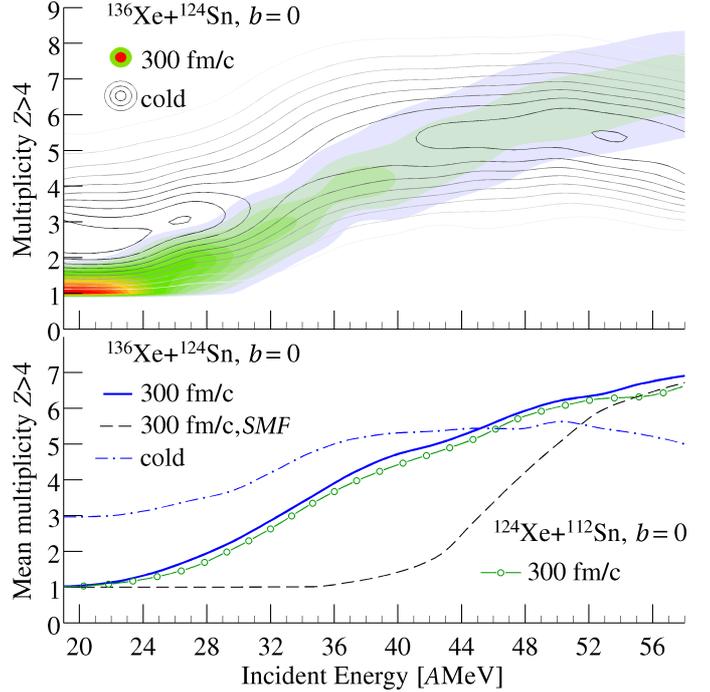}
\end{center}\caption
{
	(Color online) Top. Multiplicity distribution of fragments with $Z\!>\!4$ 
as a function of the incident energy in $^{136}$Xe$+^{124}$Sn ($b\!=\!0$) 
at $300$ fm/c (filled contours) and after secondary decay (contour lines),
	Bottom. Corresponding mean values; 
additional calculation for the system 
$^{124}$Xe$+^{112}$Sn is also shown.
	The heaviest system is also simulated within the SMF approach.
}
\label{fig_M_full}
\end{figure}
	Having verified that Langevin fluctuations show up with the expected 
properties, we let vary the incident energy over a large interval, 
from 19 to $58\,A$MeV.
	The corresponding evolution of the fragment multiplicity,
calculated by counting all fragments with charge 
number\footnote{
	Nuclear fragments are identified through a coalescence 
	algorithm in phase-space which defines the corresponding mass and charge 
	content. The fragment charge Z and mass A are approximated to integer numbers 
	under the constraint of mass, charge, momentum and energy conservation.
}
$Z\!>\!4$ 
standing out at $300$ fm/c, is shown in fig.~\ref{fig_M_full}.
	The general behaviour that BLOB traces 
is that a transition from incomplete fusion (just one nucleus observed) to multifragmentation occurs between
20 and $30\,A$MeV in $^{136}$Xe$+^{124}$Sn.
	The less neutron rich system $^{124}$Xe$+^{112}$Sn, also simulated, presents the 
same evolution, but retarded with a shift of about $2\,A$MeV to 
larger bombarding energies.
Indeed the corresponding IMF multiplicity is smaller, in agreement with experimental
findings~\cite{Moisan2012}.
	The same observable is also deduced at asymptotic time, 
by letting the system cool down after $300$ fm/c (where the excitation 
is around $3.5$MeV per nucleon) through a process of 
sequential evaporation, simulated in-flight in the swarm of light 
ejectiles and fragments (for this purpose, the model SIMON~\cite{Durand1992} was used).
%
%
%
	In the cold system the fragment multiplicity results increased by secondary decay
up to around $45\,A$MeV and levels off for larger bombarding energies due to a more
prominent decay into light fragments. 
%
%
\begin{figure}[b!]\begin{center}
	\includegraphics[angle=0, width=1\columnwidth]{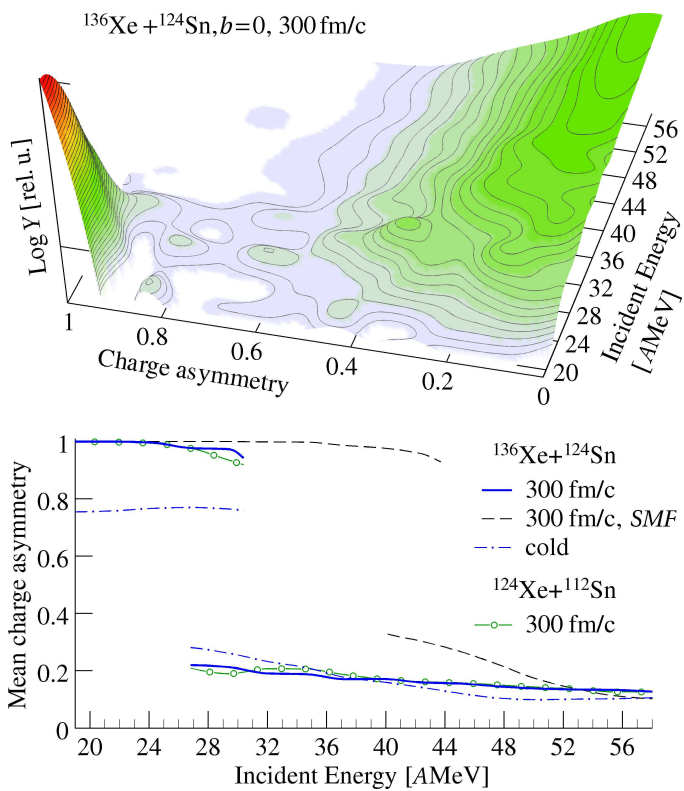}
\end{center}\caption
{
	(Color online) Top. Survey of the reaction mechanism:
distribution of charge asymmetry $\alpha$ as a function of the incident energy 
in $^{136}$Xe$+^{124}$Sn ($b\!=\!0$) 
at $300$~fm/c. 
	Bottom. Corresponding mean values.
Additional calculation for the system $^{124}$Xe$+^{112}$Sn,
	and simulation of the heaviest system within the SMF approach
and after secondary decay are also displayed.
}
\label{fig_Zasymm}
\end{figure}
%
%
\begin{figure}[tp!]\begin{center}
	\includegraphics[angle=0, width=.9\columnwidth]{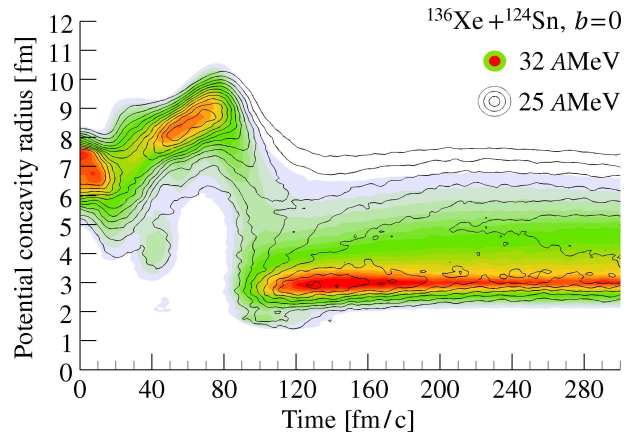}
\end{center}\caption
{
	(Color online) Temporal evolution of the size distribution of potential 
concavities in $^{136}$Xe$+^{124}$Sn ($b\!=\!0$) for an incident energy of 
$32\,A$MeV (filled contours) and $25\,A$MeV (contour lines).
}
\label{fig_concavity}
\end{figure}
	The system evolves from a dominance of incomplete fusion around $20\,A$MeV to a 
full multifragmentation pattern. 
	This result is in close accordance with 
experimental observations addressed to the same system,
where multifragmentation is already seen above $30\,A$MeV~\cite{Moisan2012}.
	On the other hand, if the BLOB approach is replaced by a 
SMF calculation with
corresponding mean-field parameters, the transition moves above $40\,A$MeV.
Hence the formation of inhomogeneities is enhanced 
in the BLOB approach
and, beside a possible additional amplification from spinodal modes, 
this process is induced by fluctuations which are 
implemented in full phase space.
	Hence BLOB is able to reproduce experimental observations related to
the onset of multifragmentation better than previous mean-field-based models.
	Moreover, fluctuations have the effect of anticipating the 
formation of fragments. 
	Indeed, the fraction of energy spent in light-particle emission reduces
in favour of a larger contribution to the developing of inhomogeneities and 
of a more explosive dynamics, which drives the separation of those 
inhomogeneities into fragments.	This should reflect also in
experimental observables like charge and velocity distributions
\cite{new}. 
	On the contrary, for small bombarding energies (below $30\,A$MeV), due to 
insufficient kinetic energy, the mean field often succeeds to 
coalesce all
the inhomogeneities 
or part of them
into a compact system.

%
\subsection{Developing of bifurcations}
	Fluctuations feed the large width of the multiplicity 
distribution above the multifragmentation threshold.
	A better insight on their distribution, and on the possible mixing
of different reaction mechanisms is provided by the charge asymmetry 
$\alpha = (Z_1-Z_2)/(Z_1+Z_2)$ between the largest fragment $Z_1$ and 
the second largest fragment $Z_2$, emitted in each event.
	This observable, 
presented in fig.~\ref{fig_Zasymm} 
illustrates that the reaction mechanism changes from incomplete 
fusion (charge asymmetry close to 
unity) to fragmentation 
(very small charge asymmetry) discontinuously, 
passing through a bombarding-energy range (around $28\,A$MeV ) where 
both mechanisms coexist.
Thus we observe, at this transition energy, a wide, but discontinuous, bunch of trajectories and
a bimodal behaviour of $\alpha$, indicating that either the system recompacts
or it breaks up into several pieces of similar size.
	Intermediate charge-asymmetry values are never populated.   
	It should be noted that this variety of configurations is related
to the same ``macroscopic'' initial conditions, i.e. the 
same beam energy and 
one unique value for the
impact parameter.  
	These features, often discussed in the context of the thermodynamics
of phase transitions and bimodal behaviour of the order parameter~\cite{Chomaz2000_2003,Bonnet2009,Pichon2006},
are observed here as a result of the fragmentation dynamics, governed by the 
spinodal decomposition mechanism.
As a consequence of the presence of fluctuations
in the dynamical trajectories,
the amount of light-particles early
emitted may vary from one event to the other, leading to energy fluctuations in the fragmenting system.  
It follows that this latter
behaves as in a
thermal bath and, at the transition energy, it oscillates between two configurations which
are energetically favourable.    
The same behaviour is also found in the corresponding SMF approach,
but at larger incident energy.
	After secondary decay, the transition pattern 
is not washed out if events producing large fission residues are removed,
and it could be searched in experimental data relative to the same systems~\cite{Moisan2012}.  
%

	To explore in finer details the developing of instabilities, the development of concavities in the mean-field potential landscape is followed in time for the incident energies $32$ and $25\,A$MeV. 
With respect to analysing the fragment features, this analysis has the advantage of applying to earlier times, before fragment formation or when fragments do not appear in the exit channel.
The distribution of the corresponding radius is shown
in fig.~\ref{fig_concavity}.
	This radius 
probes the size of forming blobs of matter which are the fragment nesting sites; 
they may eventually separate into clusters and leave the system 
(prominent mechanism at $32\,A$MeV), 
or merge into larger fragments, or fuse back together forming a 
large residue (as clearly seen at $25\,A$MeV).
	The potential concavity-radius evolves initially with the composite-system size, 
as marked by the upper branch in fig.~\ref{fig_concavity},
and suddenly drops to smaller sizes which characterise all forming droplets 
of matter within a small variance, as expected for a spinodal multifragmentation 
picture: this latter defines 
the lowest branch in fig.~\ref{fig_concavity},
and reflects in the bimodal pattern of the 
charge-asymmetry observable shown in fig.~\ref{fig_Zasymm}.

%

%
\section{Conclusions and prospects}
	In conclusion, a promising new framework of transport modelling has been realised 
by extending the Boltzmann-Vlasov formalism to include a Langevin residual term
in the evolution of
the one-body distribution function
in full phase space and adapted to fermionic systems.
	This approach reveals to keep the specific character of a one-body theory,
as far as the description of mean-field (spinodal) instabilities is concerned, and
at the same time to solve the long-standing difficulty of introducing fluctuations
of correct amplitude in a Boltzmann formalisation.
	The model, tested on 
the transition from fusion to fragmentation in central collisions at Fermi energies, 
reveals to be closer to the observation  than previous attempts to include
a Langevin term in Boltzmann theories.
Moreover, we have identified the occurrence of bifurcations 
and bimodal behaviour 
in dynamical trajectories,
linked to the fragment formation mechanism. 
At the transition energy, the system
may either recompact or split into several pieces of similar sizes.
We would like to mention that our findings do not challenge the
validity of thermodynamical analyses evidencing the occurrence of bimodal
behaviour in nuclear fragmentation~\cite{Bonnet2009,Pichon2006}. 
Indeed a large fraction of the available phase space is populated through
the spinodal decomposition mechanism~\cite{Borderie2008}, thus legitimating the use of 
thermodynamical equilibrium concepts. 
Our calculations provide a possible explanation, based on the occurrence
of dynamical instabilities, of the origin of trajectory bifurcation and 
bimodality.
We found this phenomenology in the bimodal behaviour of quantities related to fragmentation 
observables in the case of relatively low energies and 
head-on collisions.
	As bimodality has not been searched so far in this energy-centrality conditions, the present results are 
a suggestion for future experimental research.

   Finally, we stress that the results presented here may be relevant not only for nuclear fragmentation
studies, but in general for the dynamical description of quantum many-body systems.

\section{Acknowledgements}

Clarifying discussions with M.-F. Rivet, B. Borderie, J. Rizzo and F. Sebille
are gratefully acknowledged.

\end{document}